\documentclass[]{spie}

\usepackage[]{graphicx}
\usepackage[utf8]{inputenc}		
\usepackage[T1]{fontenc}		
\usepackage{xspace} 				
\usepackage{gensymb}        
\usepackage{color}

\newcommand{\um}[0]{$\mu$m\xspace}

\title{James Webb Space Telescope Optical Simulation Testbed II. Design of a Three-Lens Anastigmat Telescope Simulator}

\author{
Élodie Choquet\supit{a}, 
Olivier Levecq\supit{a,b}, 
Mamadou N'Diaye\supit{a}, 
Marshall D. Perrin\supit{a},
Rémi Soummer\supit{a}
\skiplinehalf
\supit{a} Space Telescope Science Institute, 3700 San Martin Dr, Baltimore, MD 21218 USA; \\
\supit{b} Institut d'Optique Graduate School, Palaiseau, Saint-\'Etienne and Bordeaux, France
}

\authorinfo{Correspondence should be sent to choquet@stsci.edu}
\begin{document} 
 \maketitle 

\begin{abstract}
The James Webb Space Telescope (JWST) Optical Simulation Testbed (JOST) is a tabletop experiment designed to reproduce the main aspects of wavefront sensing and control (WFS\&C) for JWST. To replicate the key optical physics of JWST's three-mirror anastigmat (TMA) design at optical wavelengths we have developed a three-lens anastigmat optical system. This design uses custom lenses (plano-convex, plano-concave, and bi-convex) with fourth-order aspheric terms on powered surfaces to deliver the equivalent image quality and sampling of JWST NIRCam at the WFS\&C wavelength (633~nm, versus JWST’s 2.12~\um). For active control, in addition to the segmented primary mirror simulator, JOST reproduces the secondary mirror alignment modes with five degrees of freedom. We present the testbed requirements and its optical and optomechanical design. We study the linearity of the main aberration modes (focus, astigmatism, coma) both as a function of field point and level of misalignments of the secondary mirror. We find that the linearity with the transmissive design is similar to what is observed with a traditional TMA design, and will allow us to develop a linear-control alignment strategy based on the multi-field methods planned for JWST. 
\end{abstract}



\section{INTRODUCTION}\label{sec:intro} 

Currently in its integration phase with a launch scheduled in 2018, the James Webb Space Telescope (JWST), successor of the Hubble Space Telescope (HST), will revolutionize astronomy. With a primary mirror of 6.5~m diameter split in 18 hexagonal segments, JWST will be the largest infrared telescope ever launched. It will carry out observations during a five to ten year mission from the visible to the mid-infrared (0.6--28 $\mu m$) with four instruments: NIRCam\cite{Rieke2011}, NIRSpec\cite{Ferruit2012}, MIRI\cite{Wright2010}, and NIRISS\cite{Doyon2012}. The science goals of JWST are to provide decisive breakthrough in four main science fields\cite{Gardner2006}: study of the first-light objects of the Universe and of the re-ionization period, analysis of the growth and morphologic evolution of the first galaxies, understanding of star and planet formation, and determination of the physical and chemical properties of planetary systems. 

To reach these goals, JWST will combine unprecedented resolution and sensitivity, with exquisite imaging capabilities across a wide field of view. The maximal telescope wavefront error after the telescope optics will be smaller than 131~nm rms across the entire field of view, and the imaging instruments NIRCam and MIRI will both achieve a Strehl ratio better than 80~\% across their respective field of view\cite{Knight2010}.

To achieve this performance, JWST segments will have to be cophased after their deployment in space. This initial commissioning will be achieved in several successive steps, from rough alignment after deployment (position of optics of $\sim1$~mm precision, orientation and inclinations of $\sim15$~arcmin) to fine sub-micron precision alignment. The NIRCam instrument serves as the principal wavefront sensor and the other instruments are also used for wide-field alignment\cite{Acton2004,Acton2012}. As part of the development and validation of wavefront sensing and control (WFS\&C) algorithms, a one-meter scale optical model of JWST was developed at Ball Aerospace (Testbed Telescope, TBT), with the same 131 degrees of freedom as JWST for the alignment: orientation, position and power of each segment of the primary mirror, and of the secondary mirror \cite{Acton2007}.

The Space Telescope Science Institute (STScI) serves as the Science and Operations Center (S\&OC) for JWST. After launch, the S\&OC will support commissioning activities and then be responsible for the operations of JWST, in particular the regular WFS\&C activities to maintain the sub-micron optical alignment quality necessary for the science goals of JWST. In this context we have designed a tabletop experiment to simulate the telescope and the NIRCam instrument with a 1:1000 scale for the primary mirror segments. The goal of the JWST Optical Simulation Testbed (JOST) is to reproduce the main WFS\&C aspects that will be involved in commissioning the telescope from coarse to fine alignment. The capabilities of JOST include phase retrieval, wavefront sensing across a large field of view equivalent to one NIRCam module, and linear wavefront control over a wide field. The testbed will also be used for independent studies and validations of WFS\&C and for studying contingency plans or possible improvements to JWST WFS\&C. In addition, the testbed will be used as part of Integration and Test (I\&T) activities ongoing at the S\&OC.

A general overview of JOST is presented in a first paper by Perrin et al. in these proceedings~\cite{Perrin2014}, hereafter ``Paper I''.
This second paper is focused on the testbed design itself (optical/optomechanical) and associated system engineering (requirements, tolerancing, error budgets, performance analysis). We also study the response of the system for linearity as a function of misalignments and field position, to verify that control similar to the Multi-Instrument Multi-Field (MIMF) alignment step for JWST will be possible with our design.

\section{DESIGN GOAL, SCOPE, AND DERIVED REQUIREMENTS} \label{sec:concept}
\subsection{Design Goal and Scope}\label{concepts}
The design goal is to deliver imaging capabilities and performance similar to the NIRCam instrument on JWST. This will enable WFS\&C studies that are relevant to JWST's operations, both during commissioning and maintenance, and to generate data directly usable in the I\&T activities of ground systems being developed at STScI. 

The scope of JOST is not to reproduce the JWST optical properties exactly, as was achieved by the TBT at Ball Aerospace\cite{Acton2007} for formal validation of the JWST optics and alignment strategies. Instead, the scope of the testbed is to simulate the most relevant aspects and physics of WFS\&C (phase retrieval, segment phasing and wide-field alignment), which  can be achieved with a simple tabletop experiment. 
The testbed design will mimic the real telescope TMA using a three-lens system (see Sec.~\ref{sec:methodo} for the choice of a transmissive design), and the JWST primary mirror (M1) simulator will include three elements:
\begin{itemize}
\item A 37-segment micro-electro-mechanical system (MEMS) deformable mirror (DM) with hexagonal segments controllable in tip, tilt and piston (Iris AO\footnote{http://wwww.Irisao.com} PTT111L). The gap size between segments is 10--12~\um ($\sim0.1$~\% of the 1.4~mm segment size, vertex to vertex) which makes them to scale relative to the actual JWST geometry. 
\item A pupil mask conjugated with the DM to define the JWST aperture shape and limit the beam to 18 active segments and introduce the central obstruction and its support structures.
\item A powered optic (Lens L1) to provide the primary mirror power since the Iris AO DM has flat segments instead of JWST's curved segments. 
\end{itemize}
The testbed will use phase retrieval algorithms to estimate the focal plane wavefront errors, as for the fine alignment of JWST. Analysis of the wavefront across a large field of view will be used to identify the system misalignments, in particular those from the secondary mirror surrogate. The testbed will use control matrices based on an optical model as compared to empirical matrices. Wavefront sensing algorithm will be validated by direct measurements with a 4D Accufiz\footnote{http://www.4dtechnology.com} Fizeau interferometer.

\subsection{JOST Requirements\label{Sec:req}}

From the top level goals as described in Paper I and summarized above, we derived ten specific requirements on the testbed design: 
\begin{itemize}
\item \textbf{REQ-1 (Wavelength):} \label{req:lambda} The system shall be designed for monochromatic operations at HeNe laser wavelength (633~nm) for simplicity of implementation. 

\item\textbf{REQ-2 (Sampling):} \label{req:sampling} The testbed shall achieve an identical sampling at HeNe as a NIRCam short-wave channel at its WFS\&C wavelength 2.12~\um. NIRCam is Nyquist-sampled at 2.0~\um, meaning that JOST shall be Nyquist-sampled at 597~nm.

\item \textbf{REQ-3 (Field of view):} \label{req:fov} The total field of view in pixels shall be equal to that of a single NIRCam module short wavelength channel (4k~$\times$~4k pixels, $\sim$2000~$\lambda/D$ at 2.12~\um).

\item \textbf{REQ-4 (Image quality):} \label{req:totalWFE} The testbed shall reproduce an image quality comparable to NIRCam at the WFS\&C wavelength with a Strehl ratio of $SR=80$~\%. At HeNe, this SR translates to a total wavefront error (WFE) of 48~nm rms. We require a total WFE lower than 80~nm rms (SR=55~\%) for the testbed, with a goal of 48~nm rms (SR=80~\%). Based on error budget calculations and tolerancing analysis (see section \ref{sec:errorbudget}) the derived requirement for the maximum optical design WFE over the field is 20~nm rms, with a goal of 10--15~nm rms. 

\item \textbf{REQ-5 (Entrance pupil size):} \label{req:pupEntrance}To facilitate manufacturing and alignment of the pupil mask at the entrance pupil, the mask diameter shall be 15--20~mm, which is about $3\times$ bigger than the conjugated exit pupil (set by the Iris AO DM dimensions, 6.1 mm vertex-to-vertex).

\item \textbf{REQ-6 (Conjugate pupil quality):} \label{req:pupQuality} The entrance pupil mask is necessary to reproduce JWST's 18-segment geometry from the 37 segment DM, and to create the central obstruction and spiders. Therefore a high-quality pupil conjugation at the segmented mirror plane is required. The pupil distortion between the entrance pupil and the DM shall be lower than 2~\% in diameter (i.e. corresponding to 10~\% of a segment size since there are five segments along the diameter). Such a distortion in the design can be absorbed by using an entrance pupil mask slightly undersized by 10~\% of a segment in diameter, i.e. loosing only 5~\% of the outermost segments. 

\item \textbf{REQ-7 (Pupil imaging):} A pupil imaging mode must be included in the design, to enable fine alignment of the two pupils, and image the exit pupil for phase retrieval measurements. The pupil imaging system must have a sampling and image quality that enable visualization of the segment gaps of the deformable mirror. Non-simultaneous pupil imaging (e.g. using the focal plane camera) is acceptable.

\item \textbf{REQ-8 (Wavefront sensing):} \label{req:4D} The design must allow an injection point for the 4D AccuFiz interferometer beam and retro-reflection by a reference surface (e.g. reference sphere), to enable direct wavefront sensing to validate phase retrieval measurements. 

\item \textbf{REQ-9 (Phase retrieval):} \label{req:clear2} The design must include a clearance of $\pm 40$~mm along the optical axis around the camera, to enable phase retrieval measurements with at least $\pm 8$ waves of defocus.

\item \textbf{REQ-10 (Opto-mechanical clearance):} \label{req:clear1} The testbed must fit on a pre-existing table of dimensions 0.9~m $\times$ 2.4~m (3' $\times$ 8'). The minimum distance between each components shall allow sufficient space for optical mounts and access to adjustment micrometers (goal of $\sim80$~mm before CAD model optimization).

\end{itemize}

Several of these requirements are related to the choice of the detector. To achieve a large field of view equivalent to a NIRCam module (REQ-3), we choose a detector head with 4096~$\times$~4096, 9~\um pixels, specifically an SBIG STX-16803 using a Kodak KAF-16803 CCD. The pixel size of the detector also sets the f-ratio of the system in the image plane to f/30.2 with respect to REQ-2, to get the same sampling as NIRCam at 2.12~\um. In addition, assuming an entrance pupil diameter of 20~mm (REQ-5), the field of view requirement (REQ-3) defines the actual field of view of JOST to $3.4\degree \times 3.4\degree$.

\section{JOST OPTICAL DESIGN} \label{sec:design}

 \subsection{Design Methodology: A Three-Lens Anastigmat Similar To JWST}\label{sec:methodo}
 
To reach the requirement of a diffraction-limited image quality over a wide field-of-view, we oriented our design to a system similar to a three-mirror-anastigmat (TMA) system. In this section we describe the analysis that led to this approach, and detail the characteristics and performance of the design. 

\begin{figure}
  \begin{center}
  \begin{tabular}{c}
  \includegraphics[width=0.9\linewidth]{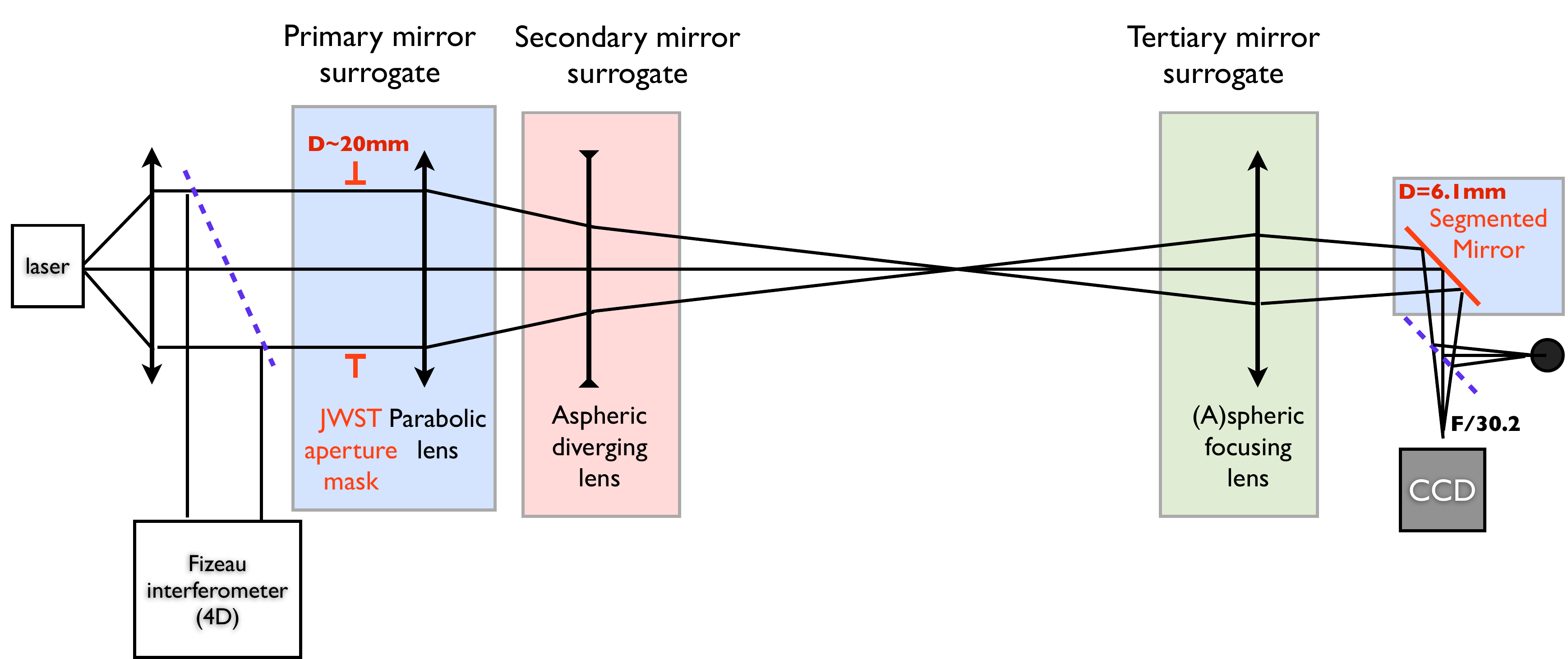}
  \end{tabular}
  \end{center}
  \caption{Optical conceptual layout of the JOST testbed, directly analogous to the optical design of the JWST optical telescope element (OTE) with the only exception that the segmented mirror is in a pupil plane distinct from the primary optics. However this conjugated pupil plane in JWST corresponds to the tip-tilt mirror used for fine pointing of the telescope\cite{knight2012} (Fast Steering Mirror), so this location is  appropriate for our segmented DM and makes our design extremely close to JWST overall. }
\label{fig:scheme}
\end{figure}

We first decided to implement a transmissive design for several reasons. With the very large field of view (REQ-3) required to simulate NIRCam-equivalent images at HeNe wavelength, a reflective TMA would have been prohibitively complex: 
\begin{itemize}
\item An on-axis TMA design requires a large entrance collimated beam, which is complex to implement for optical and mechanical reasons (requires several large high-quality optics). 
\item Off-axis reflective designs over such large field of views are challenging to design, with additional alignment complexities compared to on-axis systems. 
\item Reflective optics are not necessary for JOST since the design and operations are monochromatic (REQ-1).
\item Surface errors with refractive optics have four time less impact on the final wavefront error than reflective optics. A refractive system can thus simplify the optics specification, especially for our challenging requirement on the image quality (REQ-4).
\end{itemize}
In addition, because the M1 surrogate requires two pupil planes (the Iris AO DM and the pupil mask defining the aperture geometry and central obstruction), we found that a design directly similar to JWST \cite{knight2012} is also the most practical implementation for JOST. See Figure~\ref{fig:scheme} where M1, M2, M3 on JWST correspond to L1, L2, and L3. This layout served as the starting point for our design optimization. 

The JWST aperture surrogate consists of a pupil mask (P1) defining the entrance pupil of the system (distance P1--L1 to be optimized), and the deformable mirror in the conjugated exit pupil plane (P2) in the converging beam after L3. As with JWST there is no other pupil plane in the system, since the conjugation of P1 by L1 and L2 gives virtual images. This design has the advantage of being very similar to JWST with the Iris AO DM located at the position of the fast steering mirror (FSM) on JWST. The JWST primary mirror surrogate will also include the converging lens L1 to simulate the power of the primary mirror, initially assumed parabolic or close to parabolic similarly to JWST M1.

The JWST secondary mirror surrogate consists of a diverging lens (L2). As with the actual JWST telescope it will slow down the beam and correct for coma and astigmatism left by L1, and will produce an unused focal plane between L2 and L3. On JWST this is where a field mask, the Aft Optical System (AOS) Entrance Aperture, is located to reduce background and stray light. The L2 lens will be mounted on a tip/tilt and XYZ translation stage to simulate the five degrees of freedom for M2 active control and study its impact on the wavefront. 

The JWST tertiary mirror surrogate consists of a converging aspheric lens (L3). On JWST the M3 mirror produces a focal plane which is re-imaged by each of the science cameras. On JOST, L3  plays the role of both M3 and NIRCam, providing directly the final image on the detector with the same sampling as NIRCam at the WFS\&C wavelength. 

\subsection{JOST Wavefront Error Budget}\label{sec:errorbudget}

The total WFE error specified in REQ-4 can be split into four contributions: wavefront errors from the design itself, surface errors from the lenses, alignment errors, and Iris AO DM surface error (surface quality of 20~nm rms or possibly better). Note that the design WFE is a deterministic term and therefore directly adds to the total WFE, while other random contributions add up in quadrature.  The following error budget was therefore established to guide the overall design and achieve the 80~nm rms WFE requirement, with a goal of 50~nm rms (REQ-4):
\begin{itemize}
\item Optical design WFE: lower than 20~nm rms (goal 10~nm rms).
\item Lenses surface WFE: total lower than 30~nm rms (goal 20~nm rms), corresponding to 24~nm rms per each lens surface (goal 16~nm rms) with 6 surfaces in total (There is a factor $\sqrt{6}/2$ conversion from surface error per lens to total WFE). 
\item Alignment WFE: lower than 34~nm rms total (goal 20~nm rms). This alignment tolerance was initially verified using Zemax modeling to correspond to reasonable mechanical tolerances for alignment ($\sim 50~\mu m$).
\item Iris AO DM WFE: 40~nm rms WFE, i.e. 20~nm rms surface error or better (goal 30~nm rms). Based on vendor specification.
\end{itemize}
The surface error allocated for the lenses correspond to about 0.3~fringe peak-to-valley (PV) irregularity (i.e. $0.15~\lambda$ PV, so 24~nm rms assuming a typical factor of $\sim4$ between PV and rms). While challenging for aspheric optics this was identified to be within feasible manufacturing capabilities. Moreover, this specification applies to the entire clear aperture of the lenses. If the beams propagate through only part of the lenses (for lenses far from the pupil planes), the WFE will actually be better than the specification.

This error budget defines an extremely challenging optical design problem, with the requirement to create a three-lens anastigmat system delivering a WFE better than 20~nm rms over a $3.4\degree \times 3.4\degree$ field of view, and with a goal of 10~nm rms. Note that because of the different pixel size of our detector ($9~\mu m$ pixels) and different wavelength (633~nm vs. $2.12~\mu m$) this actual field of view is considerably larger than the actual field of view of a NIRCam module ($2.2' \times 2.2'$).

\subsection{Optical Design Optimization and Results}

\subsubsection{Three-lens anastigmat final design}
Based on this analysis and on the requirements listed in Sec.~\ref{Sec:req}, we found using Zemax\footnote{http://www.zemax.com} calculations the optimal design consisting of three custom lenses, respectively plano-convex, plano-concave and bi-convex, with a single 4th order aspherical term on each lens (L3 has one spherical surface).
In order to minimize cost, our goal was to limit the number of aspheric surfaces and aspheric terms to the minimum possible. The optimization initially involved several aspheric terms, then was simplified iteratively and restricted to the 4th order aspheric term only. Optimization was also tested with spherical surfaces for L1 or L2, but the wavefront error of the design increased significantly beyond our requirement (e.g. 21~nm rms on axis and 29~nm rms at corner of the field of view with no aspherical term for L2). This demonstrates that our system is well constrained since removing any one degree of freedom significantly degrades the performance. This is thus the optimal design within the requirements that we defined for the JOST system. 

\begin{figure}
  \begin{center}
  \begin{tabular}{c}
  \includegraphics[width=0.9\linewidth]{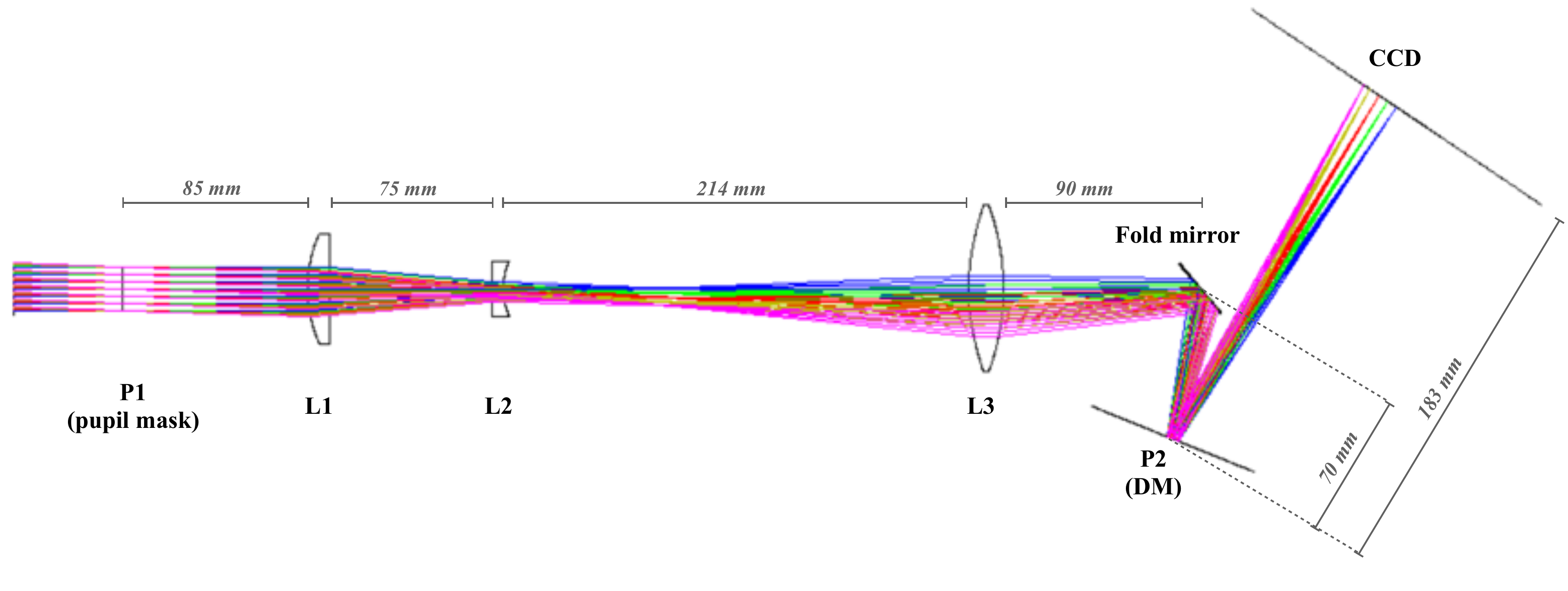}
  \end{tabular}
  \end{center}
  \caption{Final optical layout of the JOST testbed. Five different fields of view are simulated with image positions evenly spreading along the CCD diagonal, from its center (blue rays) to its right corner (purple rays). The black line at the DM position shows the clearance needed for its electronic card, and the black line at the CCD position shows the size of the detector box. \label{fig:finalScheme} }
 \end{figure}

The final optical layout of the testbed design is shown in Figure~\ref{fig:finalScheme}, and a summary of the three custom lenses' characteristics is presented in Table~\ref{tab:lenses}. 

\begin{table}
\caption{JOST lens characteristics.}
\label{tab:lenses}
\begin{center}    
\begin{tabular}{cccccc} 
Optics name 	&Type	    &Glass	&Focal length	&Diameter  &4th order aspheric coeff.\\
		  	&		    &		& (mm)		  &(mm)    &(mm$^{-3}$)\\
\hline
L1	    	&Plano-convex  &N-BK7	&119.01	  &50.8      &-2.825e-7\\
L2	    	&Plano-concave &N-BK7	&-64.45	  &25.4      &3.718e-6\\
L3	    	&Bi-convex	  &N-BK7	&103.21	  &76.2      &-5.123e-7
\end{tabular}
\end{center}
\end{table}

 The design meets or exceeds all requirements for JOST. Optimized for a wavelength of 633~nm (REQ-1), the final design entrance pupil is fixed to 20~mm diameter (REQ-5) and the f-ratio in the image plane is set to f/30.2 (REQ-2). 
The total physical size of the design is 660 by 200~mm, with a minimal distance of 70~mm between the optics, and thus meets (REQ-10). The clearance between optics has been further studied with an opto-mechanical CAD model of the testbed (see Sec.~\ref{sec:optomech}). 
The camera has about 100~mm clearance upstream and 200~mm downstream, which is enough to defocus it for phase retrieval measurements, and add a reference mirror for direct WFE measurements with the 4D interferometer (REQ-8 and REQ-9). An image of the pupil can be obtained on the detector using an additional pupil imaging lens (not described in this paper), and moving the camera to a further position (REQ-7). 

The total wavefront errors of the design are presented in Table~\ref{tab:WFE} for different field coordinates, assuming perfect alignment and perfect optical surfaces. With a maximum of 14~nm rms at the edge of the field of view and an average of 10~nm rms over the field of view, the design WFE meets the goal (10~nm rms) on average and exceeds requirement over the full field of view (REQ-3 and REQ-4).
 
\begin{table}
\caption{Wavefront errors of the final optical design for different locations within the field of view. The field coordinates are defined in the detector plane, where the on-axis field has coordinates (0, 0) at the center of the CCD of size 36~mm.}
\label{tab:WFE}
\begin{center}    
\begin{tabular}{cc} 
Field coordinates  & Simulated design WFE rms\\
(mm)        & (nm)\\
\hline
(0.0, 0.0)      &4.2  \\
(4.5, 4.5)     &5.95  \\
(9.0, 9.0)     &7.8  \\
(13.5, 13.5)    &8.2  \\ 
(18.0, 18.0)    &13.8  
\end{tabular}
\end{center}
\end{table} 

 \subsubsection{Tolerancing analysis}
Based on this final design and the complete set of specification from the selected manufacturer\footnote{http://www.optimaxsi.com}, we performed both a sensitivity analysis and a Monte Carlo tolerancing analysis  with Zemax, assuming 0.5 fringe surface irregularity ($\lambda/4$~PV), and rigid-body tolerance of $\pm 50~\mu m$ for $x,y,z$ decenters and $\pm 1.2$~ arcmin for tip/tilts. We also included the manufacturer parameters for lens surface relative decenters and tilts. This tolerancing analysis therefore includes the combined effect of the optical design itself, surface errors from lens polishing, and alignment errors (both from rigid-body positioning and lens manufacturing). 

The total WFE from sensitivity analysis is 16~nm rms and $20\pm4$~nm rms with Monte Carlo tolerancing, which are consistent. With a WFE contribution of 10~nm rms (in average over the field of view) from the optical design itself, the WFE contribution from lens surface quality and alignment combined is therefore 10~nm rms. 

We can obtain the total estimated WFE for the system by adding in quadrature the WFE from the Iris AO DM to the combined WFE from lens surfaces and alignment (10~nm rms), before adding back the deterministic contribution from the design itself. 
The total WFE is therefore $10+RSS(20-10,40)$, corresponding to a total WFE of 51~nm rms, which is better than our requirement (80~nm rms) and meets our  goal (50~nm rms). Since the actual as-built WFE of the Iris AO DM has not been evaluated at this time, this error budget includes sufficient margin to meet requirement REQ-4.
 
\subsubsection{Pupil image quality}
\begin{figure}
  \begin{center}
  \begin{tabular}{c}
  \includegraphics[width=0.9\linewidth]{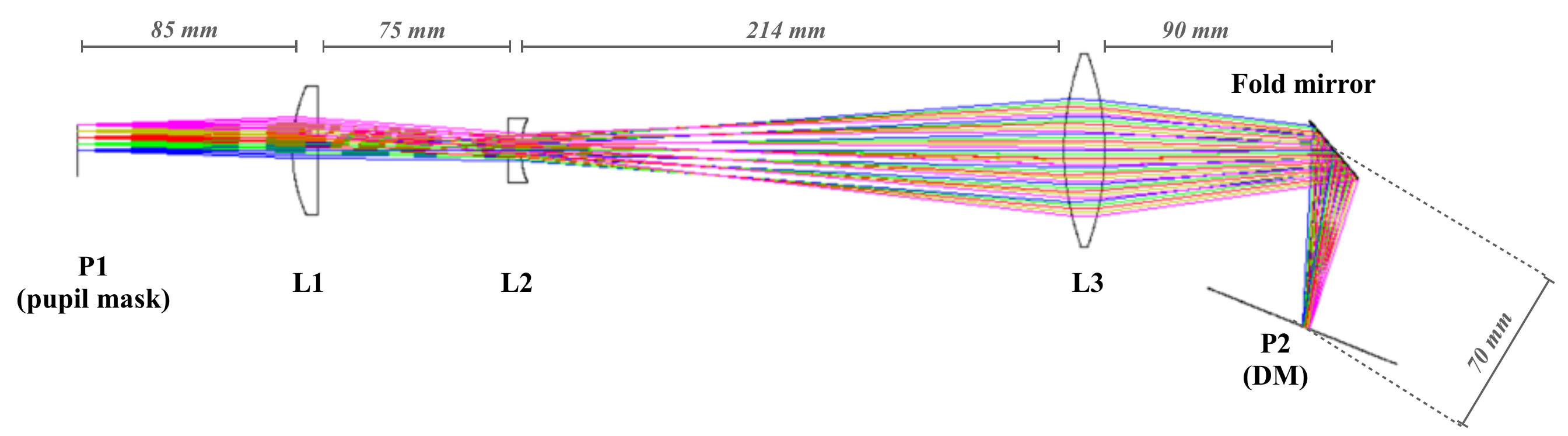}
  \end{tabular}
  \end{center}
  \caption{Image of the entrance pupil through the optical system. The beam numerical apertures are defined to propagate identical rays in the system as collimated beams with a field of view of $\pm 1.7\degree$ (corresponding to $\pm18$~mm in the focal plane). \label{fig:pupil} }
 \end{figure}

The impact on pupil quality has been evaluated by simulating a 20~mm object at the entrance pupil position, and telecentric beams with a numerical aperture corresponding to the maximal field of view in the system (see Figure~\ref{fig:pupil}). The spot sizes in the DM plane simulated by ray tracing give the maximum possible distortion and lateral defocus. We optimized the system to minimize the pupil distortion on-axis. The pupil conjugation quality is thus fully limited by the defocus induced by the angle of incidence on the DM.

The star simulator in our system will only generate one point source at a time in the entire field of view (using a single-mode fiber, an off-axis parabola, and two steering mirrors to simulate a perfect collimated beam). As a consequence, the effective numerical aperture of the beams passing through each point in the entrance pupil plane will be close to zero, unlike for a flat field (as in Fig.~\ref{fig:pupil}) or a crowded field of view. The simulated spot sizes thus capture the effect of pupil wandering and distortion combined, for all field points simultaneously, and therefore indicates the maximum amount of wandering or distortion that can exist over the field of view. The effective pupil image quality for any one given field point will thus be much better.

Table~\ref{tab:pupDist} shows the maximum pupil wander and distortion combined with respect to the angle of incidence on the DM. The optimal design with the smallest incident angle given the mechanical clearance of the system ($12\degree$ on the DM) provides a maximal pupil distortion of 2.3~\% of the DM size (including pupil wandering across the field of view). This value is slightly higher than the pupil imaging quality requirement (REQ-6) but is very conservative since it combines all field points simultaneously. In practice, if the pupil wandering or distortion turns out to be too high, we will use an entrance pupil slightly undersized to avoid light reflection on the unused outer ring of segment of the DM.

\begin{table}
\caption{Pupil wander in the DM plane with rays spanning the entire JOST field of view, for different incident angles on the segmented mirror. The pupil wander values are obtained by simulating the image of the entrance pupil through the optical system and measuring the maximum spot radius.}
\label{tab:pupDist}
\begin{center}    
\begin{tabular}{cc} 
DM incident angle  & Maximal pupil wander\\
($\degree$)     &(\um)\\
\hline
0  &  59\\
6  &  98\\
12 &  138\\  
15 &  160
\end{tabular}
\end{center}
\end{table}

 \subsection{Optomechanical Design}\label{sec:optomech}
 
\begin{figure}\begin{center}
  \begin{tabular}{c}
  \includegraphics[width=0.8\textwidth]{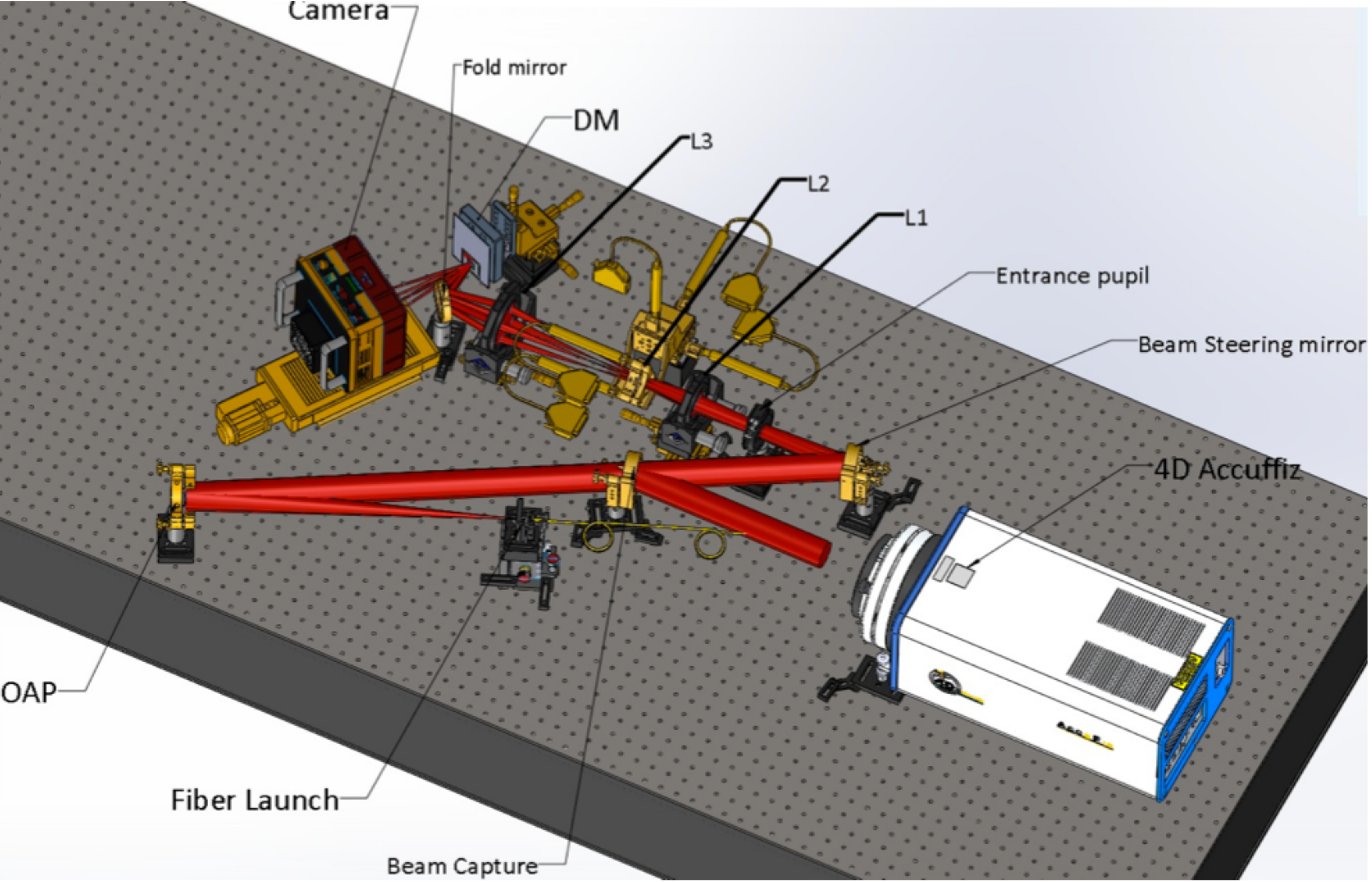}
  \end{tabular}
  \end{center}
  \caption{\label{fig:solidworks} Optomechanical design for the testbed, as rendered in Solidworks CAD software. The beam path is shown in red, including both input illumination options (the beam from the white rectangular 4D interferometer, and that from the single-mode fiber collimated into a flat wavefront using an off-axis parabola (OAP). Note the tight mechanical clearances for the CCD camera, fold mirror, DM and L3. Note also multiple beams corresponding to various field points.} 
\end{figure}

Several iterations were performed between optical optimization and opto-mechanical design using Solidworks\footnote{http://www.solidworks.com}. The most challenging part of the optomechanical design is due to the combination of the large size of the CCD package (SBIG STX-16803\footnote{http://www.sbig.com}), the translation stage for phase retrieval (REQ-9) and the Iris AO DM package and mounting (see Figure \ref{fig:solidworks}). 
 
Ideally, the DM should be used at normal incidence to optimize the conjugation with the pupil mask. This could be implemented using a beam splitter but we discarded this possibility for two reasons. 1: Using a beam splitter in a converging beam requires a very high quality optical component (face parallelism and surface quality) to limit additional wavefront errors. 2: It would generate multiple fringe patterns when using the 4D interferometer, caused by the retro-reflections between the DM and the beam splitter. We therefore accepted to use the DM at non-normal incidence, but still wish to use it as close to normal as possible. Because of mechanical clearances, an unacceptably large incidence angle would be required on the deformable mirror to accommodate the size of the L3 and DM mounts. To avoid this situation, the final design includes an additional fold mirror between L3 and the deformable mirror. This enables to minimize the incident angle on the DM while preserving sufficient mechanical clearances for the mounts and for phase retrieval. The present design thus uses the DM with a $12\degree$ incident angle, compatible with REQ-6 (conjugate pupil quality) as discussed above.  

The mechanical design also included considerations for including five motorized axes for L2 ($x,y,z$ translations and tip/tilt), with tight clearances as shown in Figure \ref{fig:solidworks}). 

The solution selected for the star simulation system involves an off-axis parabola (OAP) delivering a flat wavefront, followed by two flat mirrors (capture mirror and steering mirror). The beam at the steering mirror is sufficiently oversized (2~inch beam) compared to the entrance pupil (P1) to produce the simulated star for all field points. 

 \section{Optical properties of JOST three-lens anastigmat}

The baseline experiment for this testbed is to study the alignment state of the primary mirror segments (Iris AO DM) together with the secondary mirror surrogate (L2). We assume here that the other optics are aligned using other means (e.g. using the Fizeau interferometer or an alignment telescope).
 
An important and useful property of a TMA design is linearity of some Zernike modes both as a function of the source position in the field of view and also as a function of the magnitude of the misalignment\cite{Thompson2008,Schechter2011}. This property is particularly interesting to establish a linear-control approach for the wide-field alignment of the TMA. This is the approach adopted for JWST to sense and control the position of the secondary mirror in the Multi-Instrument Multi-Field (MIMF) stage of commissioning\cite{Acton2012a}. Since we have derived an optical design that is very similar to the TMA but in transmission, we analyzed if this property holds true for the JOST design.
Independently from the segment cophasing errors and assuming that the segmented mirror is perfectly flat, we therefore need to check that we have: 
\begin{itemize}
\item Linear dependence of the defocus and astigmatism Zernike terms as a function of field position for a given static misalignment of L2.
\item Linear dependence of the Zernike terms for a given field point as a function of the amount of misalignment of L2.
\end{itemize}

\subsection{Linearity With Misalignment Magnitude}

 \begin{figure}[p]
  \begin{center}
  \begin{tabular}{c}
  \includegraphics[width=0.95\linewidth]{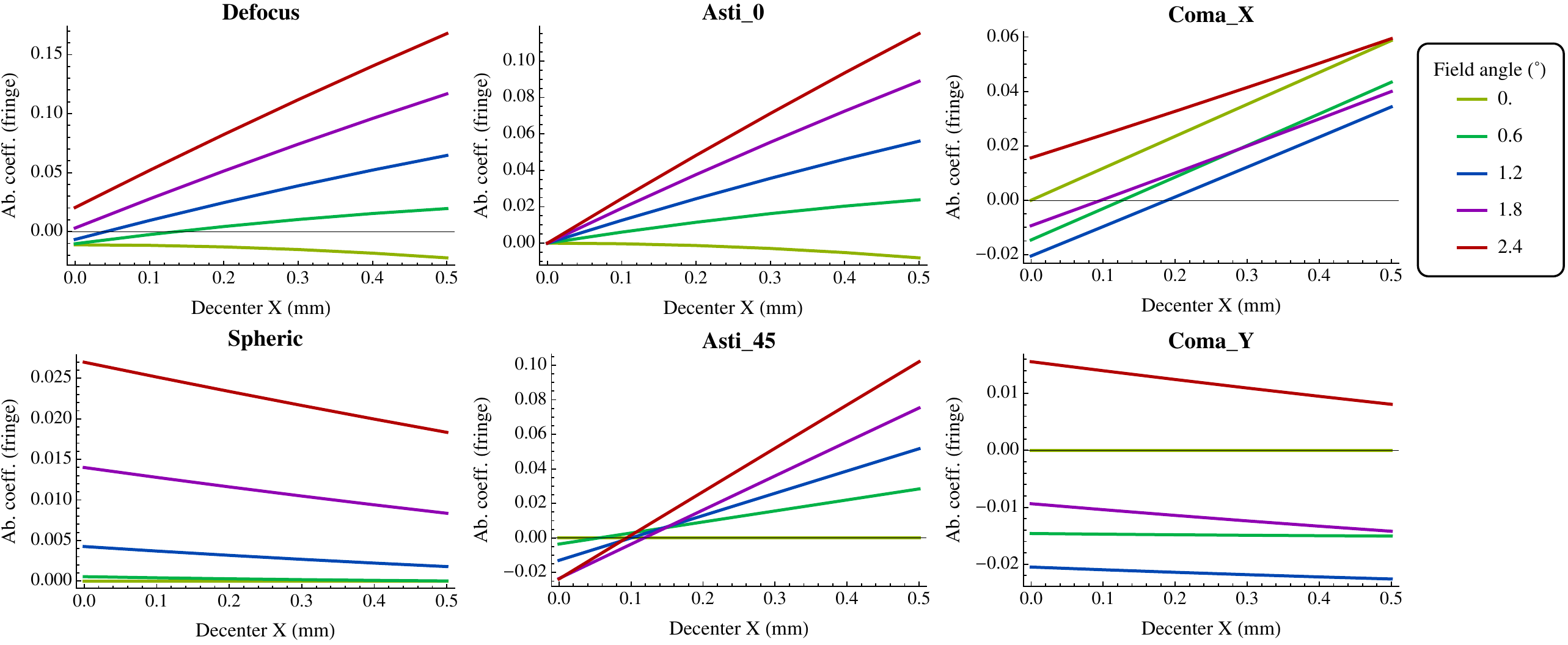}
  \end{tabular}
  \end{center}
  \caption 
  {Zernike coefficients as a function of lateral decenter of L2 along x-axis for different field angles. Note the Y axis scales differ for each plot. The astigmatism and defocus terms are the largest aberrations by a significant factor. \label{fig:misaDXvsFIELD} }
  \end{figure}

 \begin{figure}[p]
  \begin{center}
  \begin{tabular}{c}
  \includegraphics[width=0.95\linewidth]{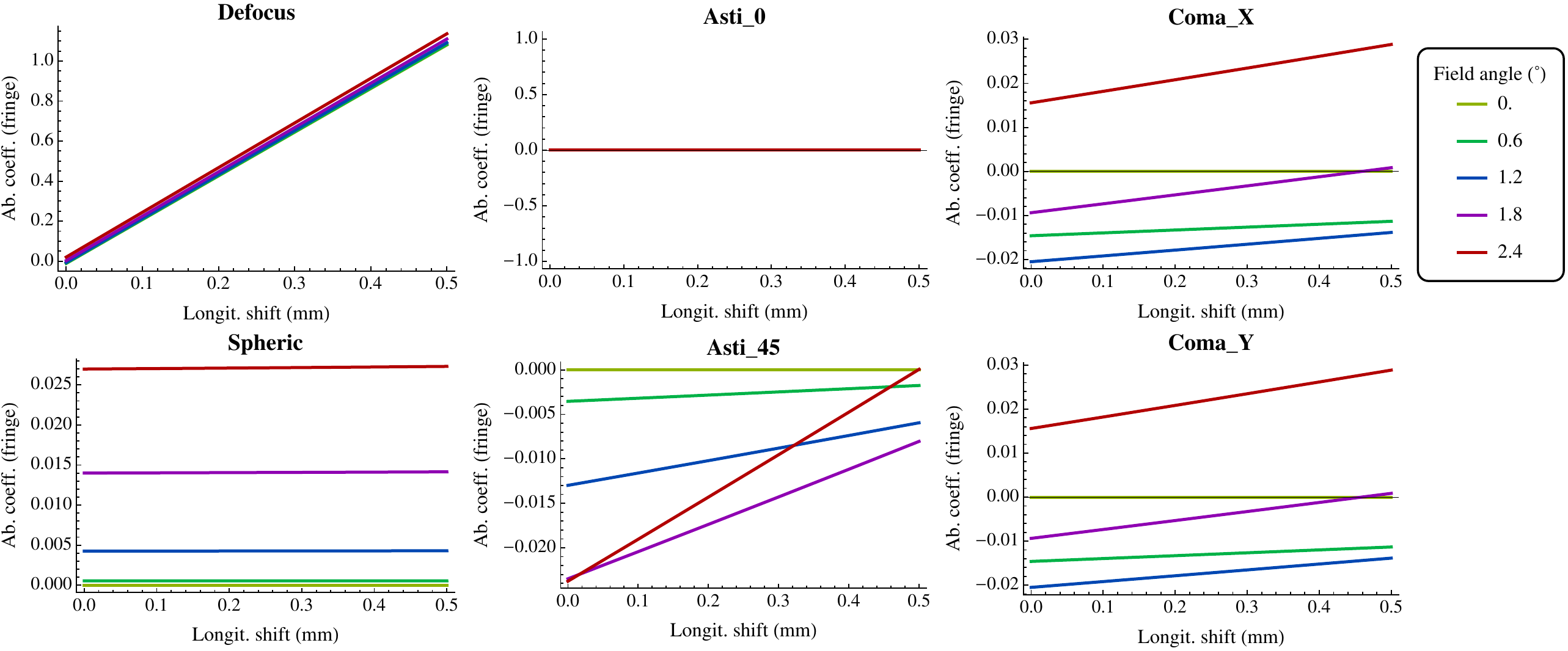}
  \end{tabular}
  \end{center}
  \caption 
  {Zernike coefficients as a function of longitudinal misalignment of L2 along optical axis, for different field angles. This is in essence a pure defocus misalignment and the Zernikes reflect that. \label{fig:misaDZvsFIELD} }
  \end{figure}

 \begin{figure}[p]
  \begin{center}
  \begin{tabular}{c}
  \includegraphics[width=0.95\linewidth]{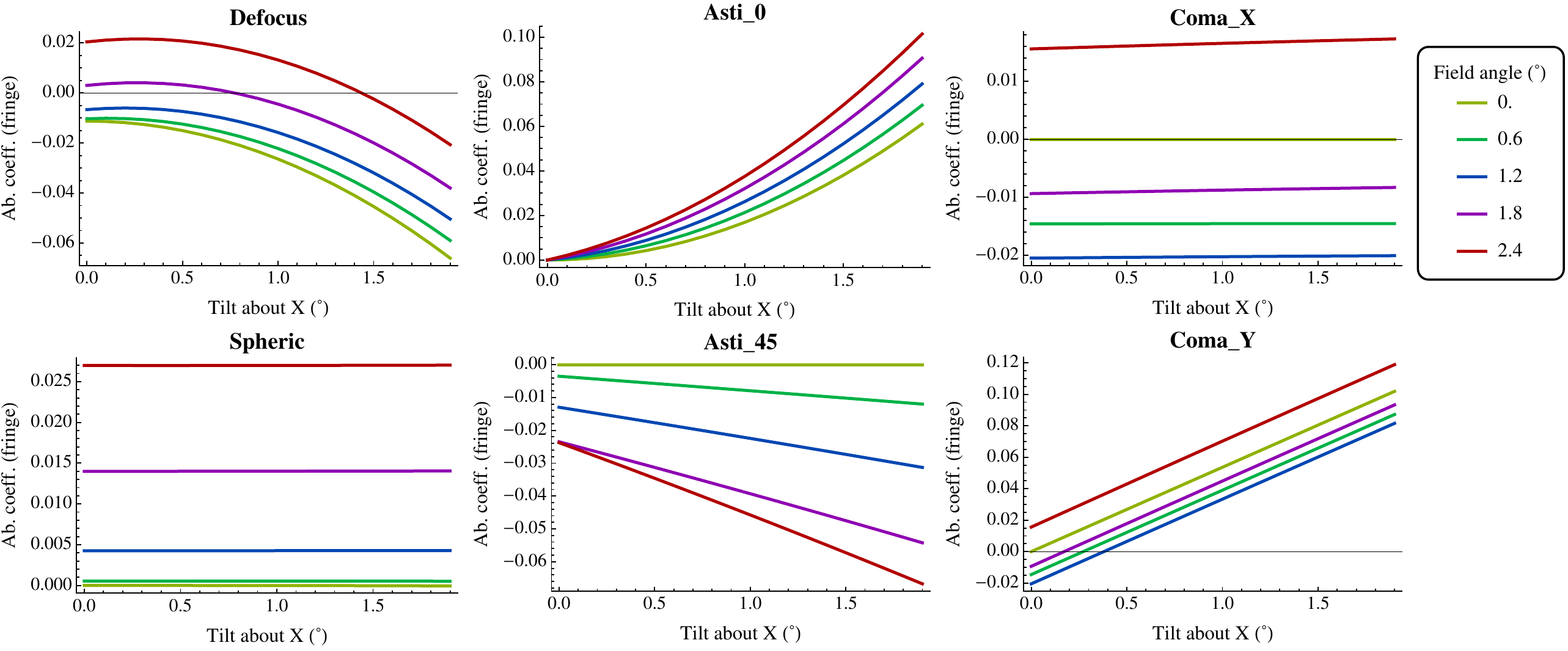}
  \end{tabular}
  \end{center}
  \caption 
  {Zernike coefficients as a function of inclination of L2 about the x-axis, for different field angles. Again note the differing Y axes scales. \label{fig:misaTXvsFIELD} }
  \end{figure}

 Figs.~\ref{fig:misaDXvsFIELD},~\ref{fig:misaDZvsFIELD}, and~\ref{fig:misaTXvsFIELD} present the six first Zernike coefficients --- Focus, spherical aberration, coma and astigmatism --- as a function of lateral decenter along the $x$-axis, longitudinal shift along the optical $z$ axis, and tilt about the x$-$axis of L2, respectively, for five field points (coordinates along the field diagonal). These plots verify the linearity of the coma, and astigmatism for these types of misalignments (x or z-axis decenter or tilting of L2), as analyzed by [\citenum{Thompson2008}]. In addition, defocus and spherical aberration are also linear with these misalignments. 

\subsection{Linearity With Field Angle}

 \begin{figure}[p]
  \begin{center}
  \begin{tabular}{c}
  \includegraphics[width=0.95\linewidth]{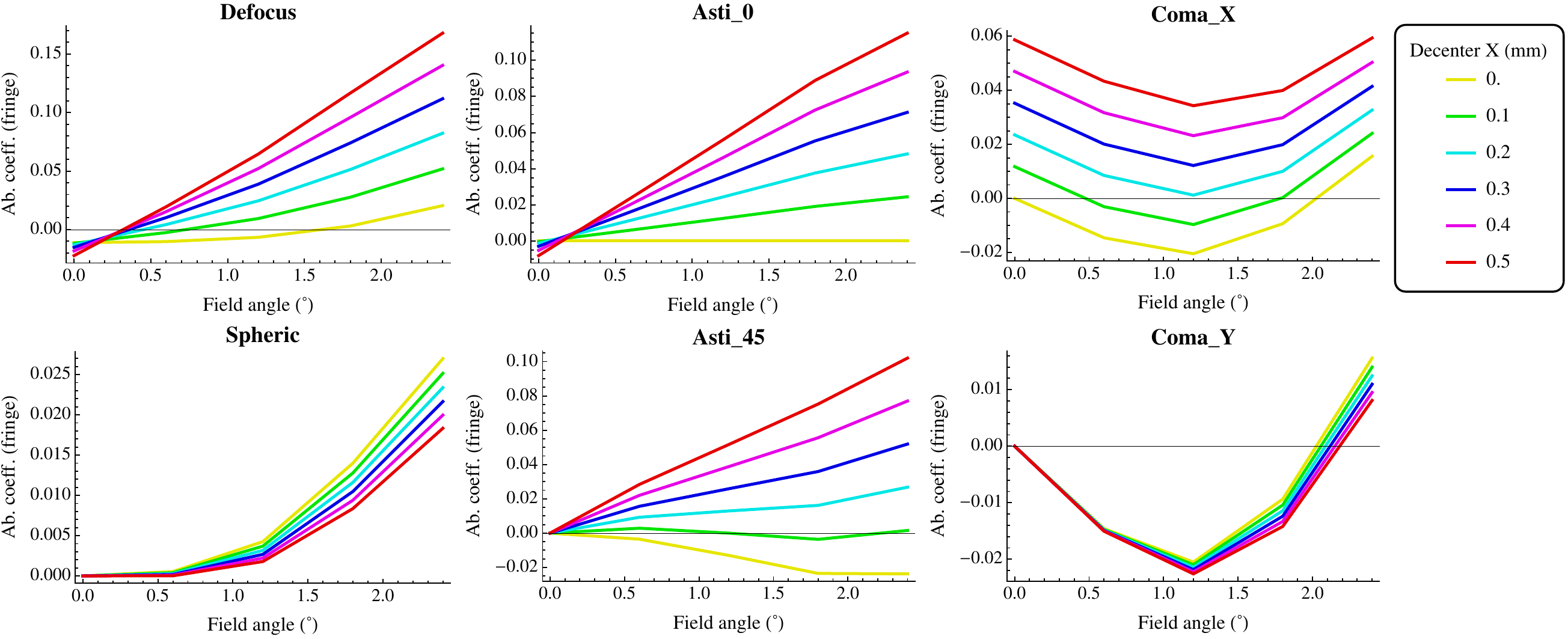}
  \end{tabular}
  \end{center}
  \caption 
  {Zernike coefficients as a function of field angle for different values of lateral decenter of L2 along x-axis. Once again the defocus and both astigmatisms are the largest terms, and are reasonably linear. \label{fig:misaFIELDvsDX} }
  \end{figure}

 \begin{figure}[p]
  \begin{center}
  \begin{tabular}{c}
  \includegraphics[width=0.95\linewidth]{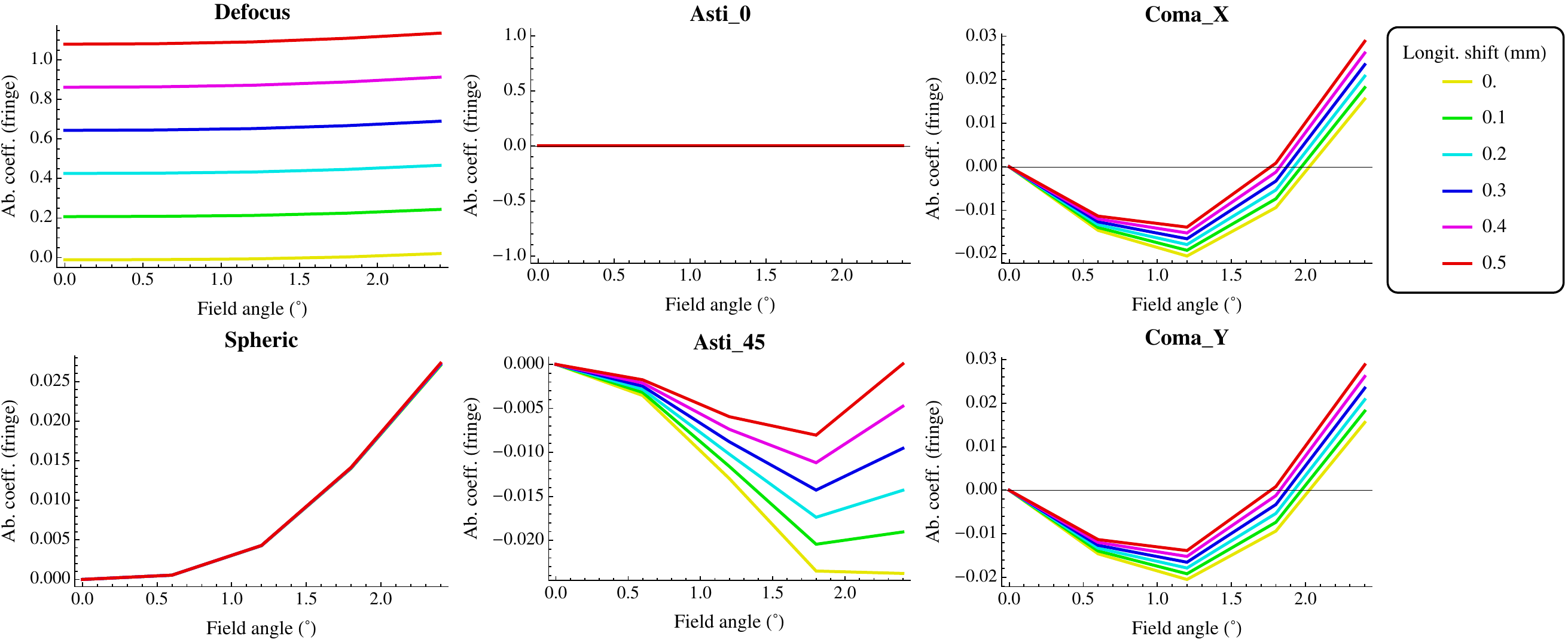}
  \end{tabular}
  \end{center}
  \caption 
  {Zernike coefficients as a function of field angle, for different values of longitudinal misalignment of L2 along optical axis.\label{fig:misaFIELDvsDZ} }
  \end{figure}

 \begin{figure}[p]
  \begin{center}
  \begin{tabular}{c}
  \includegraphics[width=0.95\linewidth]{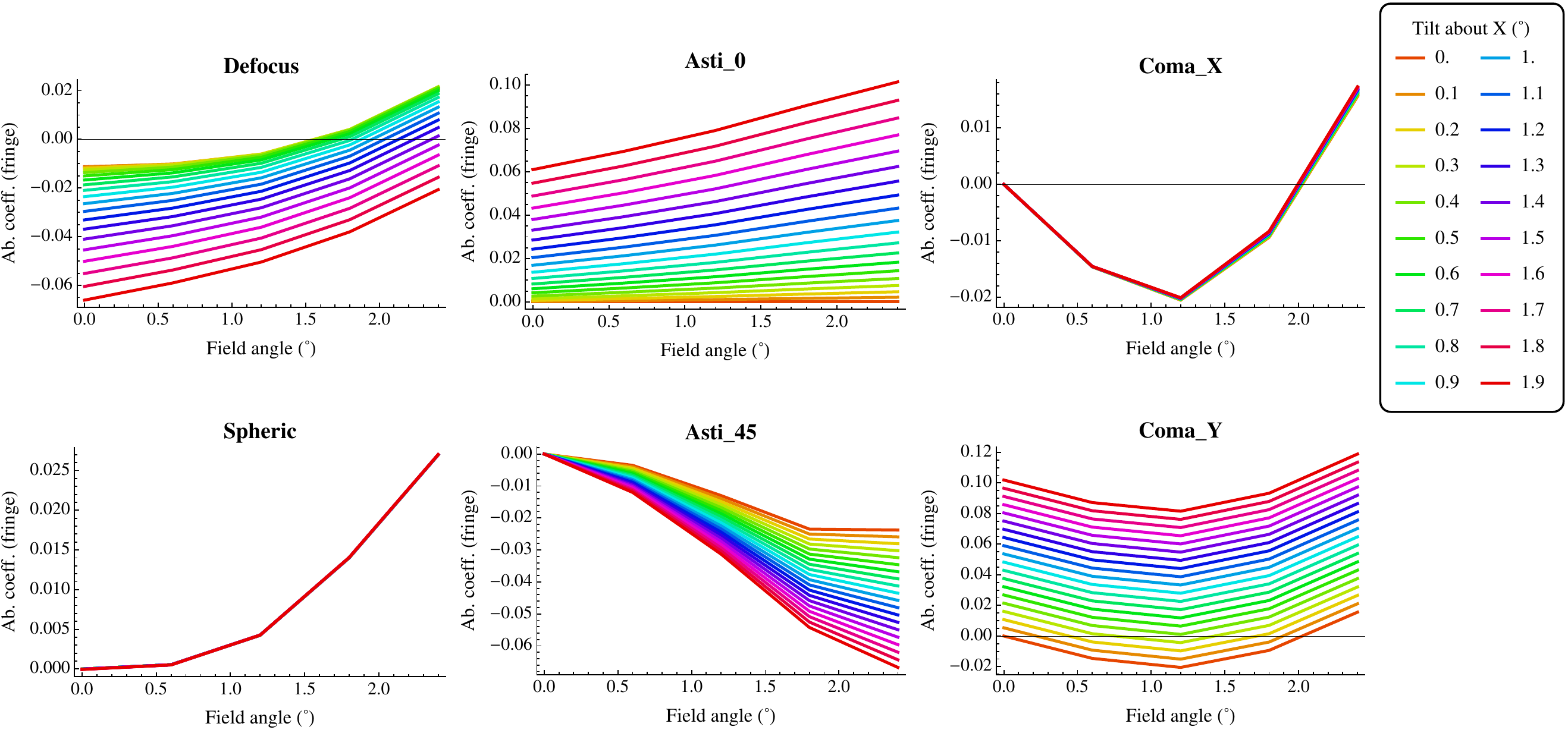}
  \end{tabular}
  \end{center}
  \caption 
  {Zernike coefficients as a function of field angle, for different inclinations of L2 about the x-axis.\label{fig:misaFIELDvsTX} }
  \end{figure}

Figs.~\ref{fig:misaFIELDvsDX},~\ref{fig:misaFIELDvsDZ} and~\ref{fig:misaFIELDvsTX} present again the same six Zernike coefficients but here as a function of the position of the source in the field of view, respectively for several amount of decenter along the X axis, defocussing along the z axis, and tilt about the x-axis of the lens L2. 
For a given misalignment type, the astigmatism is reasonably linear across the field of view as predicted by [\citenum{Thompson2008}]. The $45\degree$ astigmatism is not perfectly linear for a very wide field of view when L2 is misaligned along the optical axis.

In practice the control of the JWST M2 will be done using three terms: defocus and both astigmatisms. These terms are expected to dominate the overall wavefront error for a misaligned system. In this design, the other aberrations are generally much smaller than these three aberrations. We see approximately linear behavior for these, as expected, both as a function of the field angle and of the misalignment magnitude. For non-perfectly linear behaviors, the use of a linear optical model for control will still allow convergence toward alignment with several iterations.

 \section{Conclusion}

In this paper we presented the JOST testbed with its goals, requirements, and optimized final design. What started as a seemingly straightforward exercise to develop a small benchtop testbed for segmented mirror WFS\&C in fact required the design of a diffraction-limited optical system meeting challenging wavefront error requirements over a surprisingly large field of view. We achieved this using a three-lens anastigmat optical design inspired by and derived from the three-mirror anastigmat design of JWST itself.
An error budget and tolerancing analysis show that the final design meets the requirements with margin. The JOST testbed using a segmented deformable mirror and actively controlled secondary reproduces about half the degrees of freedom of JWST, enabling a wide range of WFS\&C experiments across a field of view equivalent to a NIRCam module.
The three-lens anastigmat solution preserves the interesting linearity properties for the main aberrations both as a function of field position and as a function of misalignment for the secondary mirror surrogate. This  will allow the implementation of similar linear  methods for wide field optical alignment as the ``multi field'' approach used on JWST. 
The custom optics and associated optomechanics as described here have now been obtained and assembled to create the JOST testbed. Initial alignments and experiments are now in progress. 
 
\acknowledgments  
R.S. and M.P. initiated and led the overall project, E.C. and O.L. conducted the optical design, O.L. and M.N. performed the optomechanical optimization. 
We are grateful to Erin Elliott, George Hartig and Laurent Pueyo for helpful advices on the design and for validating the experiment during JOST final design review.
We also thank Scott Acton, Scott Knight, Bruce Dean, Tom Zielinski, Roeland Van der Marel, and Matt Mountain for interesting discussions on the design. 
This work is supported by the JWST Telescope Scientist Investigation, NASA Grant NNX07AR82G (PI: C. Matt Mountain).

\bibliography{biblio_JOST}  
\bibliographystyle{spiebib}  

\end{document}